\documentclass[%
 reprint,
 amsmath,amssymb,
 aps,
]{revtex4-2}
\usepackage{xcolor}
\usepackage{graphicx}
\usepackage{dcolumn}
\usepackage{bm}
\usepackage{hyperref}

\begin{document}

\preprint{APS/123-QED}

\title{Inverse-Designed Time-Varying Multilayers for Programmable Frequency–Momentum Light Scattering}

\author{Mohammad M. Asgari\textsuperscript{1}}
\thanks{Email: mohammadmahdi.asgari@aalto.fi}

\author{Viktar Asadchy\textsuperscript{1}}
\thanks{Email: viktar.asadchy@aalto.fi}
\affiliation{\textsuperscript{1}Aalto University, Department of Electronics and Nanoengineering, Espoo 02150, Finland}

\begin{abstract}
Space-time-varying electromagnetic structures unlock wave-control mechanisms inaccessible to static media, including frequency conversion, harmonic generation, and joint control over frequency and momentum. Here, we introduce an inverse-design framework for multilayer space-time-periodic structures in which the spatial permittivity distribution and temporal modulation profile are optimized jointly. The proposed platform comprises multiple pixelated layers that are periodic along one or both transverse directions and harmonically modulated in time, enabling the direct synthesis of complex multichannel frequency-momentum scattering responses. We demonstrate two representative functionalities: independent control over the amplitudes and phases of three transmitted spatiotemporal channels and the generation of a frequency-momentum comb. Across these examples, we consider both continuous and binary permittivity distributions. To facilitate reproducibility and further development, we make the numerical implementation of the optimization framework freely available. These results demonstrate that inverse-designed time-varying multilayers provide a compact and versatile route toward programmable frequency-momentum multiplexing and the synthesis of space-time wave packets. The optimization code is made openly available to facilitate reproducibility and further development of the proposed inverse-design framework.

\begin{description}
\item[Keywords]
Metasurfaces, inverse design, adjoint optimization, time-modulation, frequency comb, space-time wave packets.
\end{description}
\end{abstract}

\maketitle

Time-varying electromagnetic systems offer a powerful route for controlling wave propagation beyond the limits of static media. In contrast to conventional structures, whose properties remain fixed during light--matter interaction or vary in time adiabatically, time-modulated media can exchange energy with electromagnetic fields and reshape their response in both time and frequency. This temporal degree of freedom enables wave phenomena that are difficult or impossible to achieve using spatial structuring alone, including frequency conversion, harmonic generation, parametric amplification, and magnet-less nonreciprocal propagation~\cite{morgenthaler1958velocity,hadad2015spacetime,galiffi2022photonics,pacheco2022timevarying,mikheeva2022space,zhang2021spacetimecoding,Zchut2023,taravati2022microwave,AOM_stack,asgari2024theory,wang2020nonreciprocity}.

Most studies of time-varying electromagnetic systems have so far considered canonical configurations that provide clear physical insight and allow analytical or semi-analytical treatment. Representative examples include temporal slabs and interfaces~\cite{ramaccia2020temporal,stefanini2022temporal,pacheco2020temporal,moussa2023observation}, time-varying surface impedances and metasurfaces~\cite{wang2023controlling,garg2022tmatrix,salary2018electrically,shi2016dynamic}, lumped time-varying circuit elements~\cite{ptitcyn2023tutorial,hecht2023first}, and resonant systems with relatively simple geometries, including modulated optical cavities, resonant bulk media and metasurfaces, spheres, cylinders, and their periodic arrays~\cite{minkov2017exact,stefanou2021light,ptitcyn2023floquet,schab2022scattering,asadchy2022parametric,wang2025expanding}.
These platforms have been essential for revealing the fundamental mechanisms of temporal modulation, but they explore only a limited region of the broader space--time design space.
Emerging applications increasingly require multifunctional control over many coupled spatial and temporal degrees of freedom. A prominent example is the synthesis of space--time wave packets, whose propagation properties arise from tailored correlations between their transverse-momentum and frequency content~\cite{kondakci2019optical,yessenov2022spacetime}. Space--time-varying metasurfaces offer a natural platform for generating such wave packets by jointly sculpting these two spectra, motivating design frameworks capable of controlling the full frequency--momentum response.

Related progress in static, passive structures has shown that
inverse-designed volumetric metacrystals can exploit their finite
thickness to control multiple spatial diffraction channels for
different incidence angles, polarizations, and operating frequencies
within a single device. Binary implementations of such structures have
also been fabricated and experimentally validated at millimeter-wave
frequencies~\cite{asgari2026metacrystals}. Extending this volumetric
inverse-design paradigm to time-varying media remains challenging
because spatial scattering and temporal-sideband coupling must be
treated and optimized simultaneously. Rigorous frequency-domain
methods have been developed for their forward analysis, including
multi-frequency finite-difference frequency-domain modeling of active
nanophotonic devices and rigorous space--time coupled-wave analysis of
patterned temporally modulated layers~\cite{shi2016multifrequency,inampudi2019rigorous}. These approaches enable first-principles modeling of coupled spatial and frequency harmonics but were formulated as forward solvers rather than high-dimensional inverse-design frameworks. Existing inverse-design studies generally optimize only limited space--time degrees of freedom. Sabri \textit{et al.} optimized the temporal bias waveform of a spatially fixed ITO metasurface~\cite{sabri2021broadband}, while Buddhiraju \textit{et al.} designed modulation-induced couplings in a prescribed resonator network without controlling spatial diffraction~\cite{buddhiraju2021arbitrary}. Garg \textit{et al.} optimized the temporal material response of an array of spatially fixed spherical scatterers~\cite{garg2025inverse}. More recently, Gelly \textit{et al.} jointly optimized geometrical parameters and modulation phases using the Bayesian method, but only for a single three-ridge metasurface targeting selected conversion efficiencies~\cite{gelly2026optimizing}. Thus, high-dimensional joint optimization of multilayer spatial geometry and local temporal modulation for controlling the complex amplitudes of multiple frequency--momentum channels remains largely unexplored.

In this work, we introduce an inverse-design framework for multilayer time-varying electromagnetic structures in which the spatial geometry and temporal modulation are optimized simultaneously. The platform consists of pixelated layers that are periodic along one transverse direction and harmonically modulated in time. In the most general formulation, each pixel can be assigned an independent static permittivity, modulation amplitude, and modulation phase, enabling flexible control over both momentum exchange and frequency conversion. Due to the large number of degrees of freedom in the design space, the structure can directly synthesize prescribed multichannel frequency--momentum scattering responses, including complex-valued transmission coefficients across selected Floquet channels.
We further show that this design freedom can be constrained toward more practical implementations. In particular, we consider a fabrication-oriented binary material distribution with two allowed static permittivity values, while retaining pixel-dependent modulation phases and using a shared modulation depth. This reduced parameterization preserves much of the spatiotemporal scattering functionality while moving the design closer to experimentally realizable multilayer platforms. To facilitate reproducibility and enable further exploration of space--time inverse design, we make the optimization code used in this work openly available.

\begin{figure}[t]
\centering
  \includegraphics[width=\columnwidth]{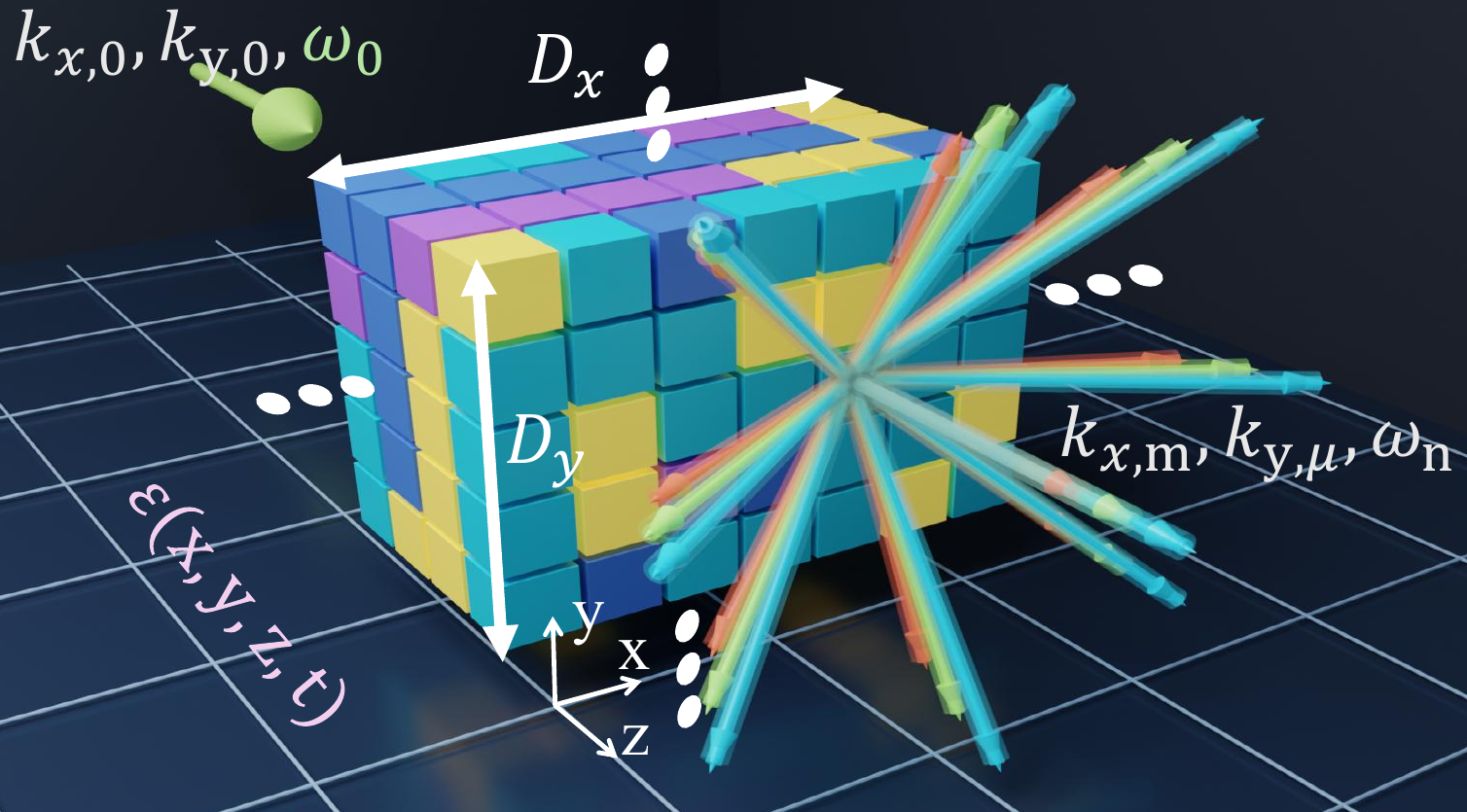}
\caption{
\textbf{Conceptual illustration of the proposed inverse-designed, time-varying electromagnetic structure.} A plane wave with frequency $\omega_0$ and transverse wave vector $(k_{x,0},k_{y,0})$ illuminates a multilayer, pixelated medium with transverse periods $D_x$ and $D_y$. In the most general framework, each pixel can have independently optimized mean permittivity, modulation depth, and modulation phase, while all pixels share the modulation frequency $\Omega$. The resulting space--time periodicity generates Floquet channels with frequencies $\omega_n=\omega_0+n\Omega$ and transverse wave-vector components $(k_{x,m},k_{y,\mu})$. Only transmitted channels are shown for clarity.
}
  \label{fig:fig1}
\end{figure}

\section{General platform and governing equations}

We consider a multilayer space--time-periodic electromagnetic structure, schematically illustrated in Fig.~\ref{fig:fig1}. The structure comprises a stack of material layers arranged along the propagation direction \(z\). Each layer is discretized into deeply subwavelength pixels in the transverse plane and may be patterned with supercell periods \(D_x\) and \(D_y\) along the transverse directions \(x\) and \(y\), respectively. The material response of each pixel is harmonically modulated in time at angular frequency \(\Omega\), common for all the pixels. The harmonic formulation below follows the extended Fourier-modal treatment of space--time-periodic media~\cite{Khorrami2021}. Although the theoretical formulation is general and allows periodicity along both the $x$- and $y$-directions, the numerical optimization examples in the next sections consider, for simplicity, structures that are periodic only along the $x-$ and uniform along the $y$-direction.

Throughout this section, \(\varepsilon\) denotes the dispersionless relative permittivity, such that the corresponding absolute permittivity is \(\varepsilon_0\varepsilon\). The relative permittivity of layer \(\ell\) is written as
\begin{equation}
\varepsilon^{(\ell)}(x,y,t)
=
\varepsilon_{\mathrm{s}}^{(\ell)}(x,y)
\left[
1+\delta^{(\ell)}(x,y)
\cos\!\left(\Omega t-\phi^{(\ell)}(x,y)\right)
\right],
\label{eq:eps_multiplicative}
\end{equation}
where \(\varepsilon_{\mathrm{s}}^{(\ell)}(x,y)\) is the static relative permittivity distribution, \(\delta^{(\ell)}(x,y)\) is the dimensionless modulation depth, and \(\phi^{(\ell)}(x,y)\) is the modulation phase.
Let \(T=2\pi/\Omega\) denote the modulation period. Because the structure is periodic in both transverse directions and in time,
\begin{equation}
\begin{aligned}
\varepsilon^{(\ell)}(x+D_x,y,t)
&=
\varepsilon^{(\ell)}(x,y,t),\\
\varepsilon^{(\ell)}(x,y+D_y,t)
&=
\varepsilon^{(\ell)}(x,y,t),\\
\varepsilon^{(\ell)}(x,y,t+T)
&=
\varepsilon^{(\ell)}(x,y,t).
\end{aligned}
\end{equation}

According to the space--time Floquet theorem, the steady-state field excited inside the time-varying material by an incident wave with transverse wave vector \((k_{x,0},k_{y,0})\) and frequency \(\omega_0\) can be written as
\begin{equation}
\mathbf{E}(x,y,z,t)
=
\exp\!\left[i(k_{x,0}x+k_{y,0}y)-i\omega_0 t\right]
\mathbf{u}(x,y,z,t),
\label{eq:floquet_form}
\end{equation}
where the Floquet envelope \(\mathbf{u}\) has the same spatial and temporal periodicities as the structure:
\begin{equation}
\begin{aligned}
\mathbf{u}(x+D_x,y,z,t)
&=
\mathbf{u}(x,y,z,t),\\
\mathbf{u}(x,y+D_y,z,t)
&=
\mathbf{u}(x,y,z,t),\\
\mathbf{u}(x,y,z,t+T)
&=
\mathbf{u}(x,y,z,t).
\end{aligned}
\end{equation}
The periodic envelope can therefore be expanded as
\begin{equation}
\begin{split}
\mathbf{u}(x,y,z,t)
&=
\sum_{m,\mu,n}
\mathbf{E}_{m,\mu,n}(z)  \\
&\quad \times
\exp\!\left[
i(mG_xx+\mu G_yy)-in\Omega t
\right],
\end{split}
\label{eq:periodic_envelope}
\end{equation}
where
\begin{equation}
G_x=\frac{2\pi}{D_x},
\qquad
G_y=\frac{2\pi}{D_y}.
\end{equation}
Substituting Eq.~\eqref{eq:periodic_envelope} into Eq.~\eqref{eq:floquet_form} gives
\begin{equation}
\begin{split}
\mathbf{E}(x,y,z,t)
&=
\sum_{m,\mu,n}
\mathbf{E}_{m,\mu,n}(z) \\
&\quad \times
\exp\!\left[
i(k_{x,m}x+k_{y,\mu}y)-i\omega_n t
\right],
\end{split}
\label{eq:field_expansion}
\end{equation}
where
\begin{equation}
k_{x,m}=k_{x,0}+mG_x,
\qquad
k_{y,\mu}=k_{y,0}+\mu G_y,
\label{eq:k_orders}
\end{equation}
and
\begin{equation}
\omega_n=\omega_0+n\Omega.
\label{eq:omega_orders}
\end{equation}
Thus, transverse spatial periodicity generates discrete diffraction orders indexed by \((m,\mu)\), whereas temporal periodicity generates discrete frequency sidebands indexed by \(n\). Each triplet \((m,\mu,n)\) defines a distinct space--time Floquet channel.

Unlike the field, the relative permittivity is strictly periodic rather than Floquet-quasiperiodic. It can therefore be expanded directly as
\begin{equation}
\begin{split}
\varepsilon^{(\ell)}(x,y,t)
&=
\sum_{r,s,q}
\varepsilon_{r,s,q}^{(\ell)} \\
&\quad \times
\exp\!\left[
i(rG_xx+sG_yy)-iq\Omega t
\right],
\end{split}
\label{eq:eps_fourier}
\end{equation}
where \(r\), \(s\), and \(q\) denote the two spatial harmonic orders and the temporal harmonic order, respectively, and all three summations run over the integers. The corresponding Fourier coefficients are
\begin{equation}
\begin{split}
\varepsilon_{r,s,q}^{(\ell)}
&=
\frac{1}{D_xD_yT}
\int_{0}^{D_x}
\int_{0}^{D_y}
\int_{0}^{T}
\varepsilon^{(\ell)}(x,y,t) \\
&\quad \times
\exp\!\left[
-i(rG_xx+sG_yy)+iq\Omega t
\right]
\,dt\,dy\,dx .
\end{split}
\label{eq:eps_fourier_coefficients}
\end{equation}

In the constitutive relation
\(\mathbf{D}=\varepsilon_0\varepsilon\mathbf{E}\),
multiplication of a permittivity harmonic \((r,s,q)\) by a field harmonic \((m,\mu,n)\) produces a field contribution with indices \((m+r,\mu+s,n+q)\). Therefore, the coefficients \(\varepsilon_{r,s,q}^{(\ell)}\) determine the coupling between the different space--time Floquet channels. Spatial patterning controls transverse-momentum exchange, whereas temporal modulation controls frequency conversion.

For the single-tone multiplicative modulation defined in Eq.~\eqref{eq:eps_multiplicative}, the permittivity contains only the temporal Fourier orders \(q=0,\pm1\). The corresponding space--time Fourier coefficients are
\begin{equation}
\varepsilon_{r,s,0}^{(\ell)}
=
\mathcal{F}_{xy,r,s}
\left\{
\varepsilon_{\mathrm{s}}^{(\ell)}(x,y)
\right\},
\label{eq:eps_q0}
\end{equation}
and
\begin{equation}
\begin{aligned}
\varepsilon_{r,s,+1}^{(\ell)}
&=
\frac{1}{2}\,
\mathcal{F}_{xy,r,s}
\left\{
\varepsilon_{\mathrm{s}}^{(\ell)}(x,y)
\delta^{(\ell)}(x,y)
e^{+i\phi^{(\ell)}(x,y)}
\right\}, \\[3pt]
\varepsilon_{r,s,-1}^{(\ell)}
&=
\frac{1}{2}\,
\mathcal{F}_{xy,r,s}
\left\{
\varepsilon_{\mathrm{s}}^{(\ell)}(x,y)
\delta^{(\ell)}(x,y)
e^{-i\phi^{(\ell)}(x,y)}
\right\}.
\end{aligned}
\label{eq:eps_qpm1}
\end{equation}
Here, \(\mathcal{F}_{xy,r,s}\{\cdot\}\) denotes the transverse Fourier coefficient evaluated over one supercell:
\begin{equation}
\begin{split}
\mathcal{F}_{xy,r,s}\{f\}
&=
\frac{1}{D_xD_y}
\int_{0}^{D_x}
\int_{0}^{D_y}
f(x,y) \\
&\quad \times
\exp\!\left[-i(rG_xx+sG_yy)\right]
\,dy\,dx .
\end{split}
\label{eq:transverse_fourier_operator}
\end{equation}

Although the permittivity contains only the temporal orders \(q=0,\pm1\), repeated coupling within the multilayer structure can populate higher-order temporal sidebands. Substituting the field and permittivity expansions into Maxwell's equations yields a coupled system for the amplitudes of all retained space--time Floquet harmonics. Following the Fourier-modal derivation~\cite{Khorrami2021}, these coupled harmonic equations can be written layer by layer for the tangential field state \(\bm{\psi}\), which includes tangential electric and magnetic fields. In our implementation, instead of multiplying transfer matrices through the stack, all layer-propagation and interface equations are assembled into a single global system
\begin{equation}
\mathbf{A}(\bm{\rho})\mathbf{x}
=
\mathbf{b},
\label{eq:global_matrix_main}
\end{equation}
where \(\bm{\rho}\) denotes the collection of design variables and
\begin{equation}
\mathbf{x}
=
\begin{bmatrix}
\mathbf{r} &
\mathbf{t} &
\bm{\psi}_0 &
\bm{\psi}_1 &
\cdots &
\bm{\psi}_{N_s}
\end{bmatrix}^{T}
\label{eq:global_unknown_main}
\end{equation}
contains the reflected amplitudes \(\mathbf{r}\), transmitted amplitudes \(\mathbf{t}\), and the tangential field states \(\bm{\psi}_j\) at the slice boundaries. The detailed assembly and boundary equations are given in Supplementary Section~1. Solving Eq.~\eqref{eq:global_matrix_main} gives the transmitted and reflected amplitudes generated from the incident channel \((m,\mu,n)=(0,0,0)\). This global-matrix form avoids long products of transfer matrices and is directly compatible with adjoint differentiation of the full scattering response, as detailed in Supplementary Section~2. The normalization of these amplitudes to the incident power flux, and the treatment of evanescent harmonics, is described in Supplementary Section~3.

The Fourier-modal implementation was successfully validated against independent reference calculations, as summarized in Supplementary Section~4. The validation includes comparison with analytical thin-film theory for a static homogeneous slab as well as COMSOL simulations of a static periodic grating, a spatially uniform time-modulated slab, and a reduced-depth pixelated spatiotemporal structure. 

\section{Inverse-design formulation}

We formulate the design problem as the optimization of a pixelated, time-modulated multilayer that maps one or more prescribed incident waves onto selected spatiotemporal scattering channels with specified complex amplitudes, including both their magnitudes and phases. Each layer is discretized into pixels along the periodic transverse directions, and the material response of each pixel is described by a static permittivity, a modulation depth, and a modulation phase. In the most general continuous parameterization, the relative permittivity of a given pixel \(i\) in layer \(\ell\) is written as
\begin{equation}
    \varepsilon_{\ell i}(t)
    =
    \varepsilon_{s,\ell i}
    \left[
    1+\delta_{\ell i}
    \cos\left(\Omega t-\phi_{\ell i}\right)
    \right],
    \label{eq:design_pixel_eps}
\end{equation}
where \(\varepsilon_{s,\ell i}\), \(\delta_{\ell i}\), and \(\phi_{\ell i}\) are design degrees of freedom. This equation is simply a discretized version of (\ref{eq:eps_multiplicative}). Depending on the physical implementation, these variables may be fully independent, shared across pixels or layers, restricted to a prescribed phase profile, or projected onto a discrete set of material values. This makes the formulation applicable both to unconstrained proof-of-principle designs and to more restricted fabrication-oriented structures.

The optimization target is defined in terms of the outgoing Floquet channels.
In the general two-dimensionally periodic formulation, a channel is labeled by
the spatial diffraction orders \(m\) and \(\mu\), the temporal sideband order
\(n\), and the polarization \(\sigma\in\{\mathrm{TE},\mathrm{TM}\}\).
Only propagating channels are included in the far-field power objectives and
reported efficiencies. A channel is propagating in exterior medium \(\nu\) when
\begin{equation}
\varepsilon_{\nu}\left(\frac{\omega_n}{c}\right)^2
-
k_{x,m}^2
-
k_{y,\mu}^2
>
0 .
\label{eq:propagating_condition}
\end{equation}
Here, $\nu\in\{\mathrm{in},\mathrm{out}\}$ denotes the homogeneous exterior medium on the input or output side of the time-modulated multilayer, respectively. A Floquet harmonic is considered propagating in a given exterior medium when it satisfies Eq.~\eqref{eq:propagating_condition}, otherwise, it is evanescent. Evanescent harmonics are retained in the Fourier-modal solution because they contribute to near-field coupling and to the electromagnetic response of the multilayer, but they are excluded from the far-field power objectives and reported channel efficiencies. All target channels considered in the numerical demonstrations below are propagating in the corresponding exterior medium.

For power-routing problems, the objective is to prescribe the power carried by a selected set of propagating Floquet channels $\mathcal{C}_{\mathrm{tar}}$. In the general two-dimensionally periodic formulation, each channel is specified by the spatial diffraction orders $m$ and $\mu$, the temporal sideband order $n$, and the polarization $\sigma$. A representative objective is
\begin{equation}
\mathcal{L}_{\mathrm{pow}}
=
\sum_{(m,\mu,n,\sigma)\in\mathcal{C}_{\mathrm{tar}}}
w_{m,\mu,n}^{\sigma}
\left(
P_{m,\mu,n}^{\sigma}
-
P_{m,\mu,n,\mathrm{tar}}^{\sigma}
\right)^2
+
\mathcal{R},
\label{eq:general_inverse_design_objective}
\end{equation}
where $P_{m,\mu,n}^{\sigma}$ is the normalized power carried by the selected Floquet channel, $P_{m,\mu,n,\mathrm{tar}}^{\sigma}$ is its prescribed target power, $w_{m,\mu,n}^{\sigma}$ is the corresponding channel weight, and $\mathcal{R}$ denotes optional regularization terms or design constraints.

The same framework can be used to prescribe complex scattering amplitudes when control over both the magnitude and phase of the output fields is required. In this case, the target is specified by the complex coefficients $T_{m,\mu,n,\mathrm{tar}}^{\sigma}$, and the objective is written as
\begin{equation}
\mathcal{L}_{\mathrm{amp}}
=
\sum_{(m,\mu,n,\sigma)\in\mathcal{C}_{\mathrm{tar}}}
w_{m,\mu,n}^{\sigma}
\left|
T_{m,\mu,n}^{\sigma}
-
T_{m,\mu,n,\mathrm{tar}}^{\sigma}
\right|^2
+
\mathcal{R}.
\label{eq:complex_amplitude_objective}
\end{equation}
Here, $T_{m,\mu,n}^{\sigma}$ denotes the power-normalized complex scattering coefficient of the selected propagating channel, which may correspond to either a reflected or a transmitted Floquet channel. With this normalization,
\begin{equation}
P_{m,\mu,n}^{\sigma}
=
\left|T_{m,\mu,n}^{\sigma}\right|^2.
\end{equation}
This objective is therefore used when both the amplitude and phase of selected output channels are prescribed.

As mentioned above, forward solution of electromagnetic scattering from the designed structure is obtained using the modified spacetime Fourier-modal solver. The optimization variables are mapped smoothly to their allowed physical ranges, and gradients are obtained by differentiating the assembled global linear system using an adjoint formulation (see Supplementary Section~2). No finite-difference gradients were used. The resulting gradient information is supplied to the Adam optimizer (adaptive moment estimation~\cite{kingma2015adam}), optionally with continuation in the number of retained spatial and temporal harmonics. This basis-continuation strategy starts the optimization in a lower-dimensional approximation of the scattering problem, which reduces computational cost and can help avoid poor local minima. The resulting design is then refined and validated with progressively larger harmonic bases.

For designs with more realistic material modulation parameters, we also
consider a binary-constrained parameterization of the static material
layout. The static permittivity is expressed in terms of a normalized
design variable \(\rho_{\ell i}\in[0,1]\) as
\begin{equation}
\varepsilon_{\mathrm{s},\ell i}(\rho_{\ell i})
=
\varepsilon_{\min}
+
\left(\varepsilon_{\max}-\varepsilon_{\min}\right)
H_{\beta}(\rho_{\ell i}),
\label{eq:binary_projection}
\end{equation}
where \(H_{\beta}\) is the smooth Heaviside projection function
\begin{equation}
H_{\beta}(\rho)
=
\frac{
\tanh(\beta\eta)
+
\tanh\!\left[\beta(\rho-\eta)\right]
}{
\tanh(\beta\eta)
+
\tanh\!\left[\beta(1-\eta)\right]
}.
\label{eq:smooth_heaviside}
\end{equation}
Here, \(\eta\) specifies the projection threshold, chosen as
\(\eta=0.5\), and \(\beta\) controls the sharpness of the projection.
During optimization, \(\beta\) is gradually increased using a
continuation scheme. In the limit \(\beta\rightarrow\infty\), the
projection approaches a step function and drives the static
permittivity toward only two allowed values \(\varepsilon_{\min}\) and
\(\varepsilon_{\max}\)~\cite{wang2011projection}.
After the static geometry is projected to a binary layout, the geometry is kept fixed while the remaining dynamical degrees of freedom, such as the modulation depth and phase, are refined. This provides a direct way to assess how much of the optimized scattering response can be retained under a simplified two-material implementation. The optimization code used to generate the designs and results presented in this work is made openly available to facilitate reproducibility and further development of the proposed framework.

\begin{figure*}[t]
\centering
\includegraphics[width=1.6\columnwidth]{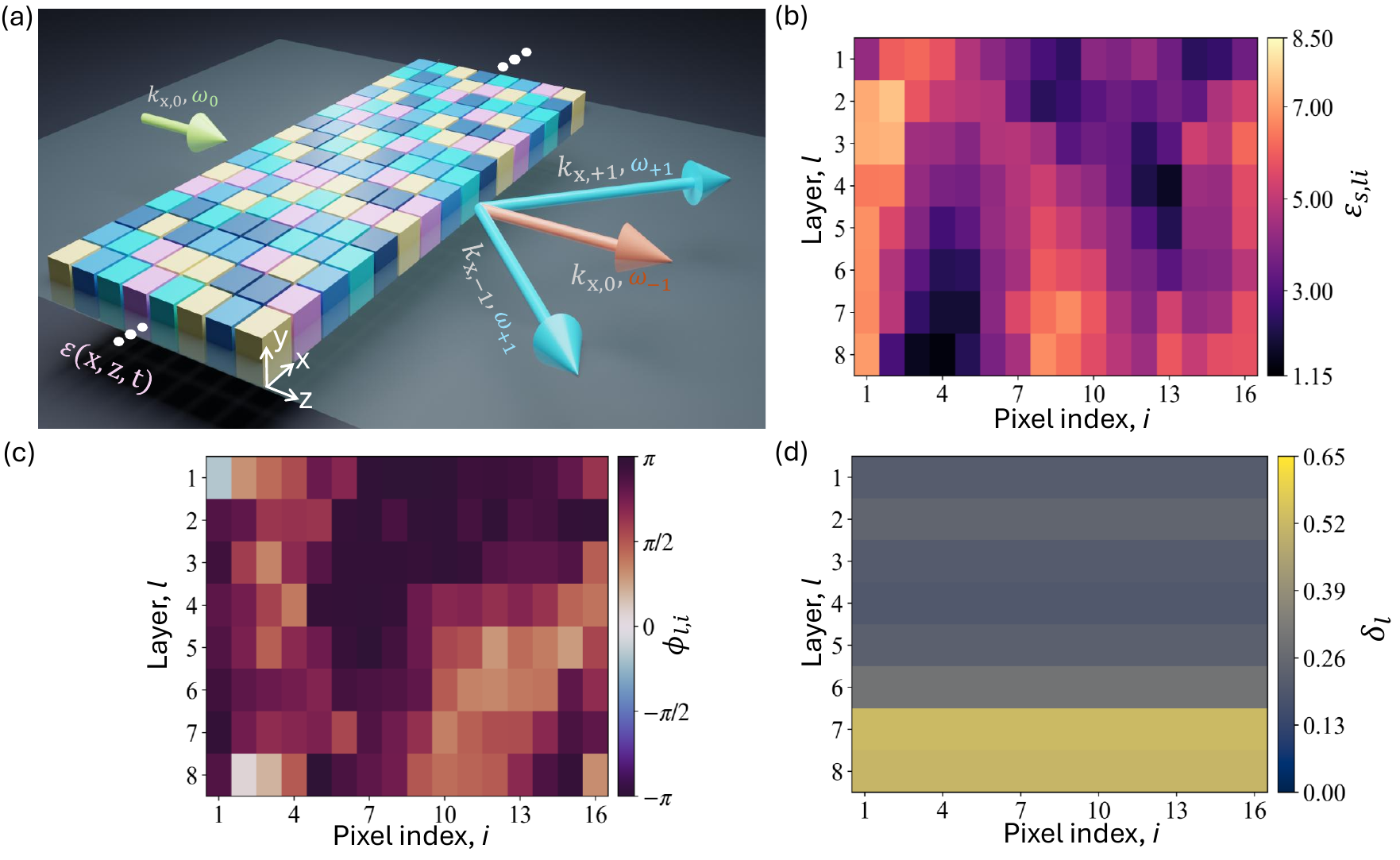}
  \caption{
\textbf{Complex-amplitude control of three independently selected spatiotemporal transmission channels.}
(a) Schematic of the optimized time-modulated multilayer. The pixel colors are assigned randomly for visualization and do not represent the optimized material parameters. A normally incident carrier \((k_{x,0},\omega_0)\) is converted into three transmitted Floquet channels with independently prescribed amplitudes and phases.  The arrow directions indicate the spatial diffraction orders, while the arrow colors indicate the generated temporal sidebands; the pixel colors in the schematic are used only for visualization.
(b) Optimized static permittivity distribution \(\varepsilon_{\mathrm{s},\ell i}\) over pixel index \(i\) and layer number \(\ell\).
(c) Optimized modulation phase distribution \(\phi_{\ell i}\).
(d) Optimized layer-dependent modulation depth \(\delta_\ell\), which is uniform along the pixels within each layer.
}
  \label{fig:fig2}
\end{figure*}

\section{Results}

To reduce the otherwise large number of accessible channels and simplify both the inverse-design procedure and the interpretation of its results, we restrict the following analysis to structures that are periodic only along the $x$-direction and uniform along the $y$-direction (only the
\(\mu=0\) subspace is retained). Consequently, diffraction occurs only along $x$, and each layer is discretized into pixels along this direction, indexed by $i$.
Therefore, in what follows, we label each far-field channel by
\((m,n,\sigma)\), where \(m\) is the \(x\)-diffraction order, \(n\) is the
temporal sideband order, and \(\sigma\in\{\mathrm{TE},\mathrm{TM}\}\) denotes
polarization. Both demonstrators consider TM-polarized excitation and target
TM-polarized output channels in transmission only. Therefore, when specifying the target channel
sets below, we suppress the polarization index and denote each target channel
simply by the pair \((m,n)\).
\subsection{Complex-amplitude synthesis in selected spatio-temporal channels}

We first demonstrate that the proposed material platform can control both the complex amplitudes and phases of selected spatiotemporal scattering channels. This is a more demanding optimization task than power routing because each prescribed complex coefficient imposes two real constraints. As mentioned above, such complex functionality is essential, for example,  for the synthesis of space-time wave packets~\cite{kondakci2019optical,yessenov2022spacetime}. 

The optimized multilayer structure is illuminated normally with a TM-polarized incident wave with the tangential wavenumber $k_{x,0}=0$ and the angular frequency $\omega_0$. 
We aim to transmit the incident wave into three spatio-temporal channels with independently and arbitrarily prescribed amplitudes and phases:
\begin{equation}
\begin{aligned}
T_{+1,+1}^{\mathrm{TM}} &\rightarrow 0.27 e^{i0^\circ},\\
T_{0,-1}^{\mathrm{TM}} &\rightarrow 0.15 e^{i100^\circ},\\
T_{-1,+1}^{\mathrm{TM}} &\rightarrow 0.45 e^{i60^\circ}.
\end{aligned}
\label{eq:three_channel_target}
\end{equation}
Apart from a weak numerical regularization applied to the outer temporal harmonics, the remaining outgoing power is allowed to be distributed among other open spatio-temporal channels.

Here \(T_{m,n}^{\mathrm{TM}}\) denotes the normalized complex transmission coefficient of the corresponding spatio-temporal Floquet channel. The normalization of the complex amplitudes and the treatment of evanescent harmonics are described in Supplementary Section~3. The geometry of the structure and the desired functionality are illustrated in Fig.~\ref{fig:fig2}(a). All the design parameters can be found in Table~I.

\begin{table}[tb]
    \centering
    \caption{Geometrical, material, modulation, and numerical parameters of the three-channel complex-amplitude demonstrator shown in Fig.~\ref{fig:fig2}. Lengths are normalized to the incident free-space wavelength \(\lambda_0=2\pi c/\omega_0\).}
\label{tab:three_channel_design_parameters}
    \begin{tabular}{l c}
        \hline
        Parameter & Value \\
        \hline
        Number of layers, \(N_L\) & 8 \\
        Pixels per layer, \(N_p\) & 16 \\
        Spatial period, \(D_x/\lambda_0\) & 1.38 \\
        Layer thickness, \(d/\lambda_0\) & 0.192 \\
        Total active thickness, \(N_Ld/\lambda_0\) & 1.54 \\
        Modulation frequency, \(\Omega/\omega_0\) & 0.15 \\
        Static permittivity range & \(1.15 \leq \varepsilon_s \leq 8.5\) \\
        Modulation depth & Layer-dependent, \(0\leq\delta_\ell\leq0.65\) \\
        Modulation phase & Pixel-dependent, \(-\pi\leq\phi_{\ell i}\leq\pi\) \\
        Validation basis & \((N_x,N_t)=(16,6)\) \\
        \hline
    \end{tabular}
\end{table}

It is worth mentioning that temporal modulation enables energy exchange between the electromagnetic field and the external modulation source. Consequently, the total outgoing electromagnetic power, defined as the sum of the reflected and transmitted powers, need not equal the incident power and may exceed it when the modulation supplies net energy to the field (it is the total photon flux which is the conserved quantity in the lossless scenario in time-varying systems~\cite{wang2021spacetime}).

We found that for the specific response given by (\ref{eq:three_channel_target}), it is enough to include  \(N_L=8\) modulated layers stacked along the \(z\)-direction, with \(N_p=16\) pixels per layer along the periodic \(x\)-direction. The pixel permittivity is parameterized as in (\ref{eq:design_pixel_eps}),

where \(\varepsilon_{s,\ell i}\) and \(\phi_{\ell i}\) are optimized for each pixel \(i\) and layer \(\ell\), while \(\delta_{\ell,i}=\delta_{\ell}\) is fixed the same for all pixels within a given layer $\ell$. 

The optimization uses the complex-amplitude objective in Eq.~\eqref{eq:complex_amplitude_objective}. For this demonstration, the target set is
\begin{equation}
\mathcal{C}_{\rm tar}
=
\{(+1,+1),(0,-1),(-1,+1)\}.
\label{eq:three_channel_target_set}
\end{equation}
We evaluated the optimized response using a \((N_x,N_t)=(16,6)\) harmonic basis. The resulting amplitudes are \(0.2698\), \(0.1438\), and \(0.4489\), in close agreement with the target values specified in Eq.~\eqref{eq:three_channel_target}. The corresponding phase errors are \(-2.18^\circ\), \(-0.37^\circ\), and \(-1.39^\circ\), respectively. Harmonic-basis tests show that these complex coefficients are numerically stable at approximately the percent level.

The three selected channels carry a total normalized transmitted power of \(0.2950\), while the remaining propagating transmitted channels carry \(0.4109\) and the reflected channels carry \(0.3922\). The resulting total electromagnetic power is \(1.0981\), which need not equal unity because energy can be exchanged with the external modulation. However, after weighting each propagating channel by \(\omega_0/\omega_n\), the total normalized photon flux summed over all transmitted and reflected channels is \(1.0000\), confirming photon-flux conservation.

The optimized static permittivity distribution, modulation phase, and layer-dependent modulation depth are shown in Figs.~\ref{fig:fig2}(b)--(d), respectively.
These results show that the inverse-designed multilayer can synthesize prescribed complex-valued scattering coefficients across independent frequency--momentum channels.

\begin{figure*}[t]
\centering
  \includegraphics[width=2.1\columnwidth]{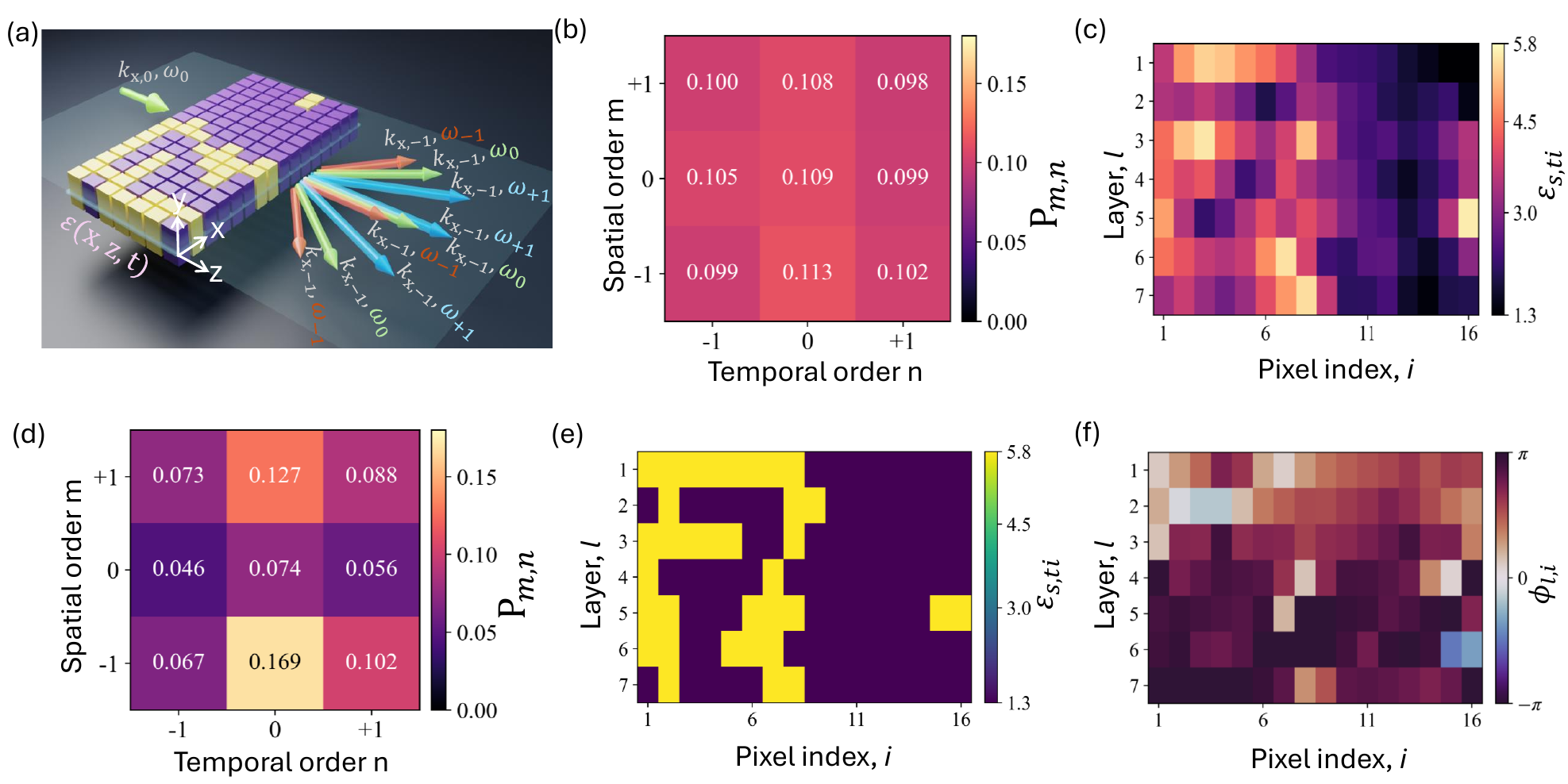}
  \caption{
  \textbf{Inverse-designed \(3\times3\) spatiotemporal comb generator.}
  (a) Schematic of the binarized time-modulated multilayer. The pixel colors are assigned randomly for visualization and do not represent the optimized material parameters.  A normally incident carrier \((k_{x,0},\omega_0)\) is converted into nine transmitted frequency--momentum channels formed by the spatial diffraction orders \(m=-1,0,+1\) and temporal sidebands \(n=-1,0,+1\).
  (b) Transmitted power distribution \(P_{m,n}\) of the optimized continuous design over the nine target comb channels.
  (c) Continuous optimized static permittivity distribution \(\varepsilon_{\mathrm{s},\ell i}\).
  (d) Transmitted power distribution \(P_{m,n}\) of the final binary design with one shared modulation depth \(\delta=0.2\), evaluated using a \((N_x,N_t)=(22,6)\) harmonic basis.
  (e) Final binary static permittivity distribution with \(\varepsilon_s\in\{1.3,5.8\}\).
  (f) Optimized modulation phase distribution \(\phi_{\ell i}\) of the binary design.
  }
  \label{fig:fig3}
\end{figure*}

\subsection{Frequency-momentum comb generator}

A key capability of a spacetime-periodic structure is the simultaneous generation of frequency sidebands and spatial diffraction orders. Temporal modulation produces sidebands at frequencies $\omega_0+n\Omega$, whereas spatial periodicity generates diffraction orders with transverse wave numbers $k_{x,0}+mG_x$. Acting together, these mechanisms produce a two-dimensional lattice of output channels in frequency–momentum space, indexed by $(m,n)$. By analogy with conventional frequency combs~\cite{fortier2019frequencycombs}, we refer to this regularly spaced set of channels as a frequency–momentum comb. In the designs considered below, we additionally seek a nearly uniform comb, in which the outgoing power is distributed approximately equally among the selected propagating channels.

To demonstrate  this functionality, we design a structure that transmits a single normally incident wave as a uniform  \(3\times3\) frequency-momentum  comb. The target channel set is
\begin{equation}
\mathcal{C}_{\mathrm{tar}}
=
\{(m,n):m,n\in\{-1,0,+1\}\}.
\label{eq:comb_target_set}
\end{equation}
The desired frequency--momentum conversion and the geometry of the optimized binary multilayer are illustrated schematically in Fig.~\ref{fig:fig3}(a). The design is optimized using the power-routing objective similar to that in Eq.~\eqref{eq:general_inverse_design_objective} with additional penalization of power coupled into non-target propagating transmission and reflection channels (see Supplementary Section~2.1). Equal target transmitted powers are assigned to the nine selected output channels, with $P_{m,n,\mathrm{tar}}^{(p)}=0.11$ for each channel. This value is chosen such that, after accounting for the different frequencies of the output channels, their total photon flux equals the incident photon flux. Because total photon flux is conserved in the lossless time-modulated system, achieving these target powers leaves no photon flux for reflected channels or undesired transmitted Floquet orders. Consequently, these unwanted channels are suppressed without requiring separate penalty terms in the objective function.

We choose the optimized structure to contain \(N_L=7\) layers stacked along the \(z\)-direction, with \(N_p=16\) pixels per layer along the periodic \(x\)-direction. All lengths are normalized to the incident free-space wavelength \(\lambda_0=2\pi c/\omega_0\). The spatial period is \(D_x=1.38\lambda_0\), each layer has thickness \(d=0.192\lambda_0\), and the modulation frequency is \(\Omega=0.15\omega_0\). The static permittivity is bounded between \(\varepsilon_s=1.3\) and \(\varepsilon_s=5.8\).

All the design parameters can be found in Table~\ref{tab:comb_design_parameters}.
The pixel-level modulation follows Eq.~(\ref{eq:design_pixel_eps}), with the static permittivity, modulation depth, and modulation phase optimized within their allowed ranges.

\begin{table}[t]
    \centering
    \caption{Design parameters of the frequency--momentum comb generator shown in Fig.~\ref{fig:fig3}. Lengths are normalized to the incident free-space wavelength \(\lambda_0=2\pi c/\omega_0\).}
\label{tab:comb_design_parameters}
    \begin{tabular}{l c}
        \hline
        Parameter & Value \\
        \hline
        Number of layers, \(N_L\) & 7 \\
        Pixels per layer, \(N_p\) & 16 \\
        Spatial period, \(D_x/\lambda_0\) & 1.38 \\
        Layer thickness, \(d/\lambda_0\) & 0.192 \\
        Total active thickness, \(N_Ld/\lambda_0\) & 1.34 \\
        Modulation frequency, \(\Omega/\omega_0\) & 0.15 \\
        Continuous permittivity range & \(1.3 \leq \varepsilon_s \leq 5.8\) \\
        Binary permittivities & \(\varepsilon_s\in\{1.3,5.8\}\) \\
        Shared modulation depth & \(\delta=0.2\) \\
        Final binary validation basis & \((N_x,N_t)=(22,6)\) \\
        \hline
    \end{tabular}
\end{table}

The powers carried by the nine target spatiotemporal channels of the optimized continuous structure are shown in Fig.~\ref{fig:fig3}(b). 
The design gives a mean target-channel power of \(0.1038\), close to the prescribed equal-energy target value \(P_{m,n,\mathrm{tar}}^{(p)}=0.11\), with a standard deviation of \(0.00512\). 
The total power in the nine selected transmitted channels is \(0.9345\), and the corresponding photon-flux sum is \(0.9492\). 
The total reflected power is only \(0.0304\). 
The continuous static-permittivity distribution of the optimized structure is shown in Fig.~\ref{fig:fig3}(c).

We further evaluate the efficiency of the frequency-momentum comb by defining the total transmitted power in the nine selected channels:
\begin{equation}
\eta_{\mathrm{comb}}
=
\sum_{(m,n)\in\mathcal{C}_{\mathrm{tar}}}
P_{m,n}^{\mathrm{TM}} .
    \label{eq:comb_efficiency}
\end{equation}

The optimized design reaches \(\eta_{\rm comb} = 0.93484\) on the
\((N_x,N_t) = (8,6)\) harmonic basis.  Re-evaluation on larger spatial bases
gives 0.93471, 0.93479, 0.93452, and 0.93450 for \(N_x = 10, 12, 14, 16\) at
\(N_t = 6\), confirming that the continuous result is well converged.

We next convert the design to a more simplified geometry with binarized static permittivity values. In this scenario, the static permittivity is restricted to two allowed material values, $\varepsilon_{s,\ell i}\in\{1.3,5.8\}$, while the variation of the modulation phase $\phi_{\ell i}$ is allowed to be continuous. To further simplify the dynamic control, the final binary design uses a single shared modulation depth for all pixels and layers,
\begin{equation}
    \delta_{\ell i} =\delta = 0.2 .
\end{equation}
The binary material response is therefore
\begin{equation}
    \varepsilon_{\ell i}(t)
    =
    \varepsilon_{s,\ell i}
    \left[
    1+\delta\cos\!\left(\Omega t-\phi_{\ell i}\right)
    \right],
    \qquad
    \varepsilon_{s,\ell i}\in\{1.3,5.8\}.
    \label{eq:binary_comb_modulation}
\end{equation}
This form separates the static binary geometry from the dynamic phase pattern: the two-material layout provides the spatial scattering structure, while the optimized phase map supplies the spacetime modulation needed to couple the target spatio-temporal channels.

The final binary shared-depth design was evaluated directly at
\((N_x,N_t) = (18,6)\), \((20,6)\), and \((22,6)\), giving
\(\eta_{\rm comb} = 0.807926\), 0.806369, and 0.803730, respectively.  At
\((22,6)\) the mean target-channel power is 0.0893, the standard deviation is
0.0357, and the reflected power is 0.1606. These results are summarized in Fig.~\ref{fig:fig3}(d). 
The optimized values of the \(\varepsilon_{s,\ell i}\) and \(\phi_{\ell i}\) are shown in Figs.~\ref{fig:fig3}(e)--(f).

The lower efficiency and larger imbalance compared with the continuous design are expected because the binary version removes intermediate permittivity values and enforces one global modulation depth. Nevertheless, most of the incident power remains concentrated in the prescribed \(3\times3\) channel set, showing that the comb-generation functionality survives under strong practical constraints.

\section{Conclusion}

We have introduced an inverse-design framework for multilayer time-varying electromagnetic structures that jointly optimizes spatial geometry and temporal modulation for desired multichannel spatiotemporal scattering. By combining pixelated spatial patterning with harmonic time modulation, the proposed platform can route an incident monochromatic wave into prescribed Floquet channels with controlled amplitudes, phases, frequencies, and transverse momenta. We demonstrated this capability through complex-amplitude synthesis in three independently selected transmitted channels and through a compact \(3\times3\) frequency--momentum comb generator. The continuous comb design directs \(93.5\%\) of the incident power into the target channel set, while a simplified binary design with a shared modulation depth retains \(80.4\%\) efficiency. These results suggest that inverse-designed time-varying multilayers can provide a versatile route toward programmable frequency--momentum multiplexing and space-time wave packets.

The results presented here can be viewed as a numerical testbed for exploring the design principles and achievable scattering functionality of time-varying multilayer systems.

Several extensions are important for moving from this idealized design setting toward experimental implementation. Material dispersion, absorption, finite modulation bandwidth, nonuniform pump delivery, and constraints on modulation phase and amplitude should be incorporated directly into the inverse-design loop. The present formulation can also be extended to include finite-size effects, multi-frequency excitation, and device-level constraints such as minimum feature size or layer connectivity. These additions would allow the same framework to optimize not only the desired spacetime scattering matrix, but also the physical feasibility of the modulation and fabrication strategy.

\vspace{2cm}
\begin{acknowledgments}
V.A. acknowledges the Finnish Foundation for Technology Promotion, and Research Council of Finland (no. 346529, 365679, and 371367). M.M.A. acknowledges a personal research grant from the Finnish Foundation for Technology Promotion (Tekniikan Edistämissäätiö, TES) for the project``Programmable Electromagnetic Materials from Near Field Holography to Wireless Communication Environments.''
\end{acknowledgments}

\bibliography{references}

\end{document}